\documentclass[twocolumn,aps,nofootinbib]{revtex4-2}
\usepackage{graphicx}
\usepackage{epstopdf}
\usepackage{hyperref}
\usepackage{latexsym}
\usepackage{amsmath}
\usepackage{amssymb}

\usepackage[utf8]{inputenc}
\hypersetup{
    colorlinks=true,
    linkcolor=red,
    citecolor=blue,
}
\usepackage{color}
\usepackage[T1]{fontenc}
\usepackage{txfonts}
\usepackage{mathrsfs}

\usepackage[center]{subfigure}

\begin{document}

 \newcommand{\bq}{\begin{equation}}
 \newcommand{\eq}{\end{equation}}
 \newcommand{\bqn}{\begin{eqnarray}}
 \newcommand{\eqn}{\end{eqnarray}}
 \newcommand{\nb}{\nonumber}
 \newcommand{\lb}{\label}
\newcommand{\PRL}{Phys. Rev. Lett.}
\newcommand{\PL}{Phys. Lett.}
\newcommand{\PR}{Phys. Rev.}
\newcommand{\CQG}{Class. Quantum Grav.}

\title{Ellis drainhole solution in Einstein-{\AE}ther gravity and the axial gravitational quasinormal modes}

\author{Kai Lin $^{a,b}$}\email{lk314159@hotmail.com}
\author{Wei-Liang Qian $^{b,c,d}$}\email{wlqian@usp.br}

\affiliation {a) Hubei Subsurface Multi-scale Imaging Key Laboratory, Institute of Geophysics and Geomatics, China University of Geosciences, Wuhan 430074, Hubei, China}
\affiliation {b) Escola de Engenharia de Lorena, Universidade de S\~ao Paulo, 12602-810, Lorena, SP, Brazil}
\affiliation {c) Faculdade de Engenharia de Guaratinguet\'a, Universidade Estadual Paulista, 12516-410, Guaratinguet\'a, SP, Brazil}
\affiliation {d) Center for Gravitation and Cosmology, College of Physical Science and Technology, Yangzhou University, Yangzhou 225009, China}

\date{Dec. 25th 2021}

\begin{abstract}
In this work, the Ellis drainhole solution is derived in Einstein-{\AE}ther gravity, and subsequently, the axial quasinormal modes of the resulting drainhole are investigated.
Owing to the presence of a minimally coupled scalar field with antiorthodox coupling polarity, the resultant metric solution is featured by a throat instead of a horizon, for which static {\ae}ther solution becomes feasible.
Moreover, the derived master equations for the axial gravitational perturbations consist of two coupled vector degrees of freedom.
By utilizing the finite difference method, the temporal profiles of the quasinormal oscillations are evaluated, and, subsequently, the complex frequencies are extracted and compared against the specific values obtained by the WKB method when the coupling is turned off.
Besides, the effect of the coupling on the low-lying quasinormal spectrum is explored, and its possible physical relevance is discussed.
\end{abstract}


\maketitle

\section{Introduction}
\renewcommand{\theequation}{1.\arabic{equation}} \setcounter{equation}{0}

The Einstein-{\AE}ther gravity originally introduced by Gasperini~\cite{agr-einstein-aether-01} is characterized by a scalar.
The latter, dictated by its dynamics, leads to the notion of a preferred state of rest at each point of spacetime.
Therefore, it is identified as the {\ae}ther field.
To be specific, the direction of the gradient of the scalar field points at the direction of time.
The norm of the gradient, on the other hand, can be interpreted as the rate of a particular cosmic clock~\cite{agr-einstein-aether-review-01}.
By stripping away the physical content associated with the norm, the {\ae}ther field proposed by Jacobson and Mattingly~\cite{agr-einstein-aether-05} further simplifies the scenario by replacing the scalar with a time-like unit vector.

An essential property of the resultant theory is that the presence of a dynamical time vector breaks the Lorentz symmetry, and in particular, it violates the boost invariance while preserving rotational symmetry in the preferred frame.
In fact, the breaking of the Lorentz symmetric is an intriguing aspect explored by a few speculations from the viewpoint of quantum gravity~\cite{agr-quantum-gravity-review-05}.
Moreover, the Einstein-{\AE}ther gravity can be viewed~\cite{agr-modified-gravity-Horava-13, agr-modified-gravity-Horava-14} as an effective low-energy theory of the Ho\v{r}ava-Lifshitz gravity~\cite{agr-modified-gravity-horava-01, agr-modified-gravity-horava-review-05}, where, in particular, the role of the {\ae}ther vector can be furnished by a khronon scalar~\cite{agr-einstein-aether-15}.
In this context, the Einstein-{\ae}ther theory serves as a consistent and concise theoretical setting to investigate the broken Lorentz invariance in the standard relativistic framework.

Theories with broken Lorentz invariance lead to various pertinent features as well as challenges in various aspects such as waveforms, stars, black holes, and cosmology~\cite{agr-einstein-aether-10, agr-einstein-aether-08, agr-einstein-aether-09, agr-einstein-aether-20}.
Static solutions resembling stars have been established~\cite{agr-einstein-aether-08}.
However, it turned out not feasible to reconcile the Killing horizon to static {\ae}ther field~\cite{agr-einstein-aether-09}.
As a result, the {\ae}ther field in such solutions always flows into the Killing horizon.
Moreover, in such theories, Hawking's rigidity theorem becomes irrelevant due to the presence of superluminal particles.
In other words, the event horizon of a stationary, asymptotically flat black hole spacetime cannot be defined in terms of the Killing horizon, which leads further to the notion of universal horizon~\cite{agr-einstein-aether-13, agr-modified-gravity-horava-lw-06}.
Subsequently, the boundary conditions of the associated black hole quasinormal modes become a complicated problem.
Nonetheless, the above complications do not significantly impact the wormhole solution.

Indeed, by itself, the wormhole~\cite{agr-wormhole-Ellis-01, agr-wormhole-Ellis-02, agr-wormhole-01, agr-wormhole-02}, as a hypothetical astrophysical object that provides a shortcut between two distant spacetime regions, is a relevant topic in the domains of classical and quantum gravity~\cite{book-wormhole-Visser}.
For instance, it is particularly interesting when a traversable wormhole solution connects the two branes~\cite{agr-extra-dim-RS-08} in the context of the Randall-Sundrum models~\cite{agr-extra-dim-RS-01, agr-extra-dim-RS-02}.
Moreover, the qusinormal modes in the wormholes, as a mean to investigate the stability of the spacetime metric in question, become a pertinent topic and have been explored by various authors~\cite{agr-wormhole-qnm-01, agr-wormhole-qnm-02, agr-wormhole-qnm-03, agr-wormhole-qnm-05, agr-wormhole-qnm-13, agr-wormhole-qnm-15, agr-wormhole-qnm-18, agr-wormhole-qnm-17, agr-wormhole-qnm-16}.
The quasinormal ringings in the wormhole to monopole perturbations were first investigated by Konoplya and Molina~\cite{agr-wormhole-qnm-01}.
The numerically extracted quasinormal frequencies are consistent with those obtained by the WKB and other methods, while the late-time tail follows a power-law, the exponent is different from that of the Schwarzchild black hole.
More recently, the gravitational wave echoes~\cite{agr-qnm-echoes-01, agr-qnm-echoes-15, agr-qnm-echoes-20} were also explored in terms of dissipative oscillations in wormholes~\cite{agr-qnm-echoes-16, agr-qnm-echoes-21}.
Following this line of thought, in the present study, we derive an Ellis drainhole solution~\cite{agr-wormhole-Ellis-01, agr-wormhole-Ellis-02} in the Einstein-{\AE}ther gravity and explore the axial gravitational quasinormal modes.

The remainder of the paper is organized as follows.
In the following section, we derive the Ellis drainhole solution and discuss its properties.
The obtained metric is static, spherically symmetric, composed of two asymptotically flat spacetime regions connected by a traversable drainhole.
The master equations of the axial perturbations are obtained in Sec.~\ref{section3}, which turn out to be a pair of coupled equations between two degrees of freedom.
The numerical approaches are carried out in Sec.~\ref{section4}.
By using the finite difference method, the temporal profiles of the quasinormal oscillations are evaluated.
The corresponding complex frequencies are extracted by the Prony method and shown to be consistent by comparing against the values obtained using the WKB method for particular cases where the couple in the master equations are switched off.
The approach is then generalized to deal with the coupled master equations, and the effect of the coupling between the two degrees of freedom in the master equations is investigated.
Further discussions and the concluding remarks are given in the last section.

\section{Drainhole solution in Einstein-{\AE}ther Gravity} \label{section2}
\renewcommand{\theequation}{2.\arabic{equation}} \setcounter{equation}{0}

We consider the Einstein-{\AE}ther theory~\cite{agr-einstein-aether-05} augmented by the inclusion of an exotic scalar field $\Phi$, which is minimally coupled to the geometry featuring antiorthodox coupling polarity~\cite{agr-wormhole-Ellis-01, agr-wormhole-Ellis-02}.
The resulting action reads
\bqn
\lb{Action1}
S&=&\frac{1}{16\pi G_{\ae}}\int\sqrt{-g}d^4x\left[R+2g^{\alpha\beta}D_\alpha\Phi D_\beta\Phi\right.\nb\\
&&-\left(c_1g^{\alpha\beta}g_{\mu\nu}+c_2\delta^\alpha_\mu\delta^\beta_\nu+c_3\delta^\alpha_\nu\delta^\beta_\mu-c_4u^\alpha u^\beta g_{\mu\nu}\right)\nb\\
&&\left.\times\left(D_\alpha u^\mu\right)\left(D_\beta u^\nu\right)+\lambda\left(u^\rho u_\rho+1\right)\right],
\eqn
where $u^\mu$ is the {\ae}ther field, it is unit time-like vector, which is enforced by the Lagrange multiplier $\lambda$.
$G_{\ae}$ is related to the Newtonian constant. 
The term composed of four dimensionless parameters $c_i$ ($i=1, 2, 3$, and $4$) of the Einstein-{\AE}ther theory~\cite{agr-einstein-aether-05} breaks the boost invariance.
The relevant range of these parameters is governed mainly by the observations and physical considerations~\cite{agr-einstein-aether-review-01}.

The resulting field equations are
\bqn
\lb{FieldEqu1}
R_{\mu\nu}-\frac{1}{2}g_{\mu\nu}R-S_{\mu\nu}&=&T_{\mu\nu}\nb\\
\text{\AE}_\mu&=&0\nb\\
g^{\alpha\beta}D_\alpha D_\beta\Phi&=&0\nb\\
g_{\alpha\beta}u^\alpha u^\beta&=&-1,
\eqn
where
\bqn
\lb{FieldEqu2}
S_{\alpha\beta}&\equiv&D_{\mu}\left[J^\mu_{(\alpha}u_{\beta)}+J_{(\alpha\beta)}u^\mu-u_{(\beta}J_{\alpha)}^\mu\right]\nb\\
&&+c_1\left[\left(D_\alpha u_\mu\right)\left(D_\beta u^\mu\right)-\left(D_\mu u_\alpha\right)\left(D^\mu u_\beta\right)\right]\nb\\
&&+c_4a_\alpha a_\beta+\lambda u_\alpha u_\beta-\frac{1}{2}g_{\alpha\beta}J^\mu_\nu D_\mu u^\nu.\nb\\
\text{\AE}_\mu&\equiv&D_\nu J^\nu_\mu+c_4 a_\nu D_\mu u^\nu+\lambda u_\mu,\nb\\
T_{\mu\nu}&\equiv&-2\left(D_\mu\Phi D_\nu\Phi-\frac{1}{2}g_{\mu\nu}D^\alpha\Phi D_\alpha \Phi\right) ,
\eqn
and
\bqn
\lb{FieldEqu3}
J^\alpha_\mu&\equiv&\left(c_1g^{\alpha\beta}g_{\mu\nu}+c_2\delta^\alpha_\mu\delta^\beta_\nu+c_3\delta^\alpha_\nu\delta^\beta_\mu\right.\nb\\
&&\left.-c_4u^\alpha u^\beta g_{\mu\nu}\right)D_{\beta}u^\nu\nb\\
a^\mu&\equiv&u^\alpha D_\alpha u^\mu .
\eqn
From above field equations, we also have
\bqn
\lb{FieldEqu4}
\lambda=u_\beta D_\alpha J^{\alpha\beta}+c_4a_\rho a^\rho .
\eqn

In the literature, it is sometimes convenient to define $c_{14}\equiv c_1+c_4$, $c_\pm\equiv c_1\pm c_3$, and $c_{123}=c_1+c_2+c_3$ in the place of the four constants $c_i$.
Another relevant choice is $\{c_S,c_V,c_T,c_\phi\}$ defined as
\bqn
\lb{cij}
c_S^2&=&\frac{c_{123}(2-c_{14})}{c_{14}(1-c_{+})(2+c_{+}+3c_2)} \nb\\
c_V^2&=&\frac{2c_1-c_{+}(2c_1-c_{+})}{2c_{14}(1-c_{+})}\nb\\
c_T^2&=&\frac{1}{1-c_{+}}\nb\\
c_\phi^2&=&\frac{c_{123}}{c_{14}} ,
\eqn
which can be interpreted, respectively, as the velocities of the scalar, vector, tensor, and khronon scalar modes~\cite{agr-einstein-aether-10, agr-einstein-aether-25}.
By inverting the above relations, one finds
\bqn
\lb{cijA}
c_{123}&=&\frac{2c_T^4c_\phi^2-2c_S^2}{c_T^4+3c_S^2c_T^2}\nb\\
c_{14}&=&\frac{2c_T^4c_\phi^2-2c_S^2}{c_T^2c_\phi^2(3c_S^2+c_T^2)}\nb\\
c_{+}&=&1-c_T^{-2}\nb\\
c_{-}&=&4c_V^2\frac{c_T^4c_\phi^2-c_S^2}{c_T^2c_\phi^2(3c_S^2+c_T^2)}-(c_T^2-1) .
\eqn
These notations will be referred to indiscriminately in the remainder of the present paper.

To proceed, we consider the following spherically symmetric metric for a static star
\bqn
\lb{metric1}
ds^2&=&-f(r)dt^2+\frac{dr^2}{f(r)}+\rho^2\left(d\theta^2+\sin^2\theta d\varphi^2\right),\nb\\
u_\mu&=&\sqrt{f(r)}\delta^t_\mu\nb\\
\Phi&=&\Phi(r) .
\eqn
By substituting above ansatz into the field equations, one encounters the following Ellis drainhole solution in Einstein-{\AE}ther gravity
\bqn
\lb{EillsSolution}
\Phi(r)&=&\sqrt{\frac{n^2-\frac{c_{14}}{2}m^2}{n^2-m^2}}\left[\frac{\pi}{2}-\text{arctan}\left(\frac{r-m}{\sqrt{n^2-m^2}}\right)\right],\nb\\
f(r)&=&\exp\left[-m\frac{\pi-2\text{arctan}\left(\frac{r-m}{\sqrt{n^2-m^2}}\right)}{\sqrt{n^2-m^2}}\right]\nb\\
\rho^2&=&\frac{r^2+n^2-2mr}{f(r)} .
\eqn
The above solution contains two positive parameters, $m$ and $n$, satisfying $n > m$.
In the place of a horizon, the resulting metric is featured by a drainhole at $r = 2m$.
There, the radius of the two-sphere $\rho$ attains the minimal value, determined by $n$.
The spherically symmetric solution is static because it admits a time-like Killing vector $\delta^t_\mu$ and is invariant under the time reflection.
Furthermore, the {\ae}ther field $u_\mu$ is also static~\cite{agr-einstein-aether-08}, as it is manifestly aligned with the above time-like Killing vector.
The obtained spacetime comprises two asymptotically flat regions joined at the drainhole.
The latter is traversable from either direction and moreover, the spacetime is geodesically complete and does not contain any singularity~\cite{agr-wormhole-Ellis-01}.
In particular, when $m=0$ and $n \ne 0$, the spacetime falls back to a nongravitating, purely geometric, Lorentz invariant, and traversable wormhole.


\section{Axial gravitational perturbations in Ellis drainhole spacetime} \label{section3}
\renewcommand{\theequation}{3.\arabic{equation}} \setcounter{equation}{0}

This section investigates the quasinormal modes of the axial gravitational perturbations in the obtained Ellis drainhole metric.
As a comparison, for a black hole solution in asymptotically flat spacetime, the notion of quasinormal modes are defined between the horizon and spatial infinity, where the ingoing and outgoing boundary conditions are introduced~\cite{agr-qnm-review-02, agr-qnm-review-03}.
For the present case, however, as the drainhole is traversable, the appropriate boundary conditions should be defined at both asymptotical spatial infinity $r\to \pm\infty$.
As discussed below, since the resultant effective potential possesses a maximum, the WKB approximation~\cite{agr-qnm-WKB-01, agr-qnm-WKB-02, agr-qnm-WKB-03} is a feasible method.
Moreover, the finite difference method~\cite{agr-qnm-finite-difference-01} can also be applied for the entire range of the radial coordinate $-\infty < r  <\infty$.

For the axial gravitational perturbations, one considers the Regge-Wheeler gauge
\bqn
\lb{metric3}
\delta g_{\mu\nu}&=&
\begin{bmatrix}
0 & 0 & 0 & h_0(r) \\
0 & 0 & 0 & h_1(r) \\
0 & 0 & 0 & 0  \\
h_0(r) & h_1(r) & 0 & 0
\end{bmatrix}
e^{-i\omega t}\sin\theta\partial_\theta P_L(\cos\theta)\nb\\
\delta u_\mu&=&\delta^\varphi_\mu h_n(r)e^{-i\omega t}\sin\theta\partial_\theta P_L(\cos\theta)
\eqn
where the method of separation of variables is adopted and it suffices to consider the case of vanishing magnetic quantum number $M=0$~\cite{agr-qnm-review-02}.
The freedom between different components with parity $(-1)^{L+1}$ is fixed by the particular choice of the gauge vector~\cite{agr-qnm-01}, and one results in two independent vector degrees of freedom in terms of $h_1$ and $h_n$.
Also, the backreactions are ignored as the perturbations are assumed to be insignificant when compared to the background, namely, $g_{\mu\nu}\gg\delta g_{\mu\nu}$ and $u_{\mu}\gg\delta u_{\mu}$.

To derive the master eqution, one further introduce the following transformation
\bqn
\lb{transformation}
h_1(r)&=&\frac{i\omega\sqrt{n^2+r^2-2mr}}{f(r)^{3/2}e^{\frac{3m\pi}{2\sqrt{n^2-m^2}}}}R_B(r)\nb\\
h_n(r)&=&R_C(r) ,
\eqn
which leads to, after some algebra, the following two coupled master equations
\bqn
\lb{masterequation1}
&&f(r)\frac{d}{dr}\left(f(r)\frac{dR_B(r)}{dr}\right)\nb\\
&&~~~~~+\left(\frac{\omega^2}{c_T^2}-V_T(r)\right)R_B(r)=U_T(r)R_C(r)\nb\\
&&f(r)\frac{d}{dr}\left(f(r)\frac{dR_C(r)}{dr}\right)\nb\\
&&~~~~~+\left(\frac{\omega^2}{c_V^2}-V_V(r)\right)R_C(r)=U_V(r)R_B(r) ,
\eqn
where
\bqn
\lb{masterequation2}
V_T(r)&=&\frac{f(r)^2}{\left(r^2+n^2-2mr\right)^2}\left[12m^2+L(L+1)r^2\right.\nb\\
&&\left.+(L^2+L-3)n^2-2(L^2+L+3)mr\right]\nb\\
U_T(r)&=&e^{-\frac{3\pi m}{2\sqrt{n^2-m^2}}}\frac{2c_{14}m f(r)^2}{\left(r^2+n^2-2mr\right)^{3/2}}\nb\\
V_V(r)&=&\frac{f(r)^2(n^2+r^2-2mr)^{-2}}{(c_-+c_+-c_-c_+)}\left\{4c_{14}^2m^2\right.\nb\\
&&+4c_{14}m(r-3m)+(c_++c_--c_+c_-)\nb\\
&&\times\left[5m^2-2(L^2+L+1)mr\right.\nb\\
&&\left.\left.+L(L+1)(r^2+n^2)\right]\right\}\nb\\
U_V(r)&=&\frac{2c_{14}m (L^2+L-2)e^{-\frac{11\pi m}{2\sqrt{n^2-m^2}}}f(r)^{-2}}{\left(c_++c_--c_-c_+\right)\left(r^2+n^2-2mr\right)^{3/2}} .\nb\\
\eqn
It is noted that the resulting system of master equations describes two coupled degrees of freedom.
In practice, the coupling poses a difficulty to most available methods to straightforwardly evaluate the quasinormal frequencies.

By inspecting Eqs.~(\ref{masterequation2}), one observes that the two potentials $U_T$ and $U_V$ identically vanish when $c_{14}=0$ or $m=0$, and subsequently, the two equations become independent for this particular case.
Moreover, the forms of the remaining potentials $V_T$ and $V_V$ are also significantly simplified, as they do not explicitly depend on $c_i$ in the decoupled master equations.
Therefore, this is the case when most conventional approaches, such as WKB approximation, can be employed to evaluate the quasinormal modes.

In what follows, we explore further Eqs.~\eqref{masterequation1} by considering two cases, $m\ne 0$ and $m=0$, separately.
For $m=0$, the master equations are readily decoupled, and one may further introduce the transformation
\bqn
\lb{rescale2}
r\rightarrow &&\ nr\nb\\
\omega\rightarrow&& \ \omega/n\nb,
\eqn
so that Eqs.~\eqref{masterequation1} become independent of $n$, which read
\bqn
\lb{masterequation3}
&&\frac{d^2R_B(r)}{dr^2}+\left(\frac{\omega^2}{c_T^2}-\frac{3-L(L+1)(1+r^2)}{(1+r^2)^2}\right)R_B(r)=0\nb\\
&&\frac{d^2R_C(r)}{dr^2}+\left(\frac{\omega^2}{c_V^2}+\frac{L(L+1)}{1+r^2}\right)R_C(r)=0 .
\eqn

On the other hand, when $m\ne 0$, it is observed that one can simplify the above equations by rescaling the coordinates and parameters using
\bqn
\lb{rescale1}
r\rightarrow&& \ mr \nb\\
\omega\rightarrow&& \ \omega/m \nb\\
n\rightarrow&& \ mn ,\nb
\eqn
so that the resultant equations does not explicitly depend on $m$.
In order words, without loss of generality, it suffices to choose $m=1$.

Moreover, in order to facilitate the numerical calculations, in the case of $m=1$, we assume that both modes propagate at the same speed, namely, $c_V=c_T$ (thus $c_-=\frac{c_+-2c_{14}}{c_+-1}$).
Subsequently, Eqs.~(\ref{masterequation2}) are simlified to read
\bqn
\lb{masterequation4}
f(r)&=&e^{-\frac{\pi-2\text{arctan}\left(\frac{r-1}{\sqrt{n^2-1}}\right)}{\sqrt{n^2-1}}},\nb\\
V_T(r)&=&e^{-2\frac{\pi-2\text{arctan}\left(\frac{r-1}{\sqrt{n^2-1}}\right)}{\sqrt{n^2-1}}}\nb\\
&&\times\frac{n^2(L^2+L-3)+(r-2)(L^2r+Lr-6)}{(n^2+(r-2)r)^2},\nb\\
U_T(r)&=&\frac{2c_{14}e^{-\frac{\pi-8\text{arctan}\left(\frac{r-1}{\sqrt{n^2-1}}\right)}{2\sqrt{n^2-1}}}}{(n^2+(r-2)r)^{3/2}},\nb\\
V_V(r)&=&e^{-2\frac{\pi+2\text{arctan}\left(\frac{r-1}{\sqrt{n^2-1}}\right)}{\sqrt{n^2-1}}}\nb\\
&&\times\frac{2c_{14}-1+(L+L^2)(n^2+(r-2)r)}{(n^2+(r-2)r)^2},\nb\\
U_V(r)&=&\frac{(L+2)(L-1)e^{-\frac{7\pi+8\text{arctan}\left(\frac{r-1}{\sqrt{n^2-1}}\right)}{2\sqrt{n^2-1}}}}{(n^2+(r-2)r)^{3/2}} .\nb\\
\eqn
The resulting potentials are illustrated in Fig.~\ref{Fig4}, which are governed by three parameters: $c_{14}$, $n$, and $L$.

Since all the potentials given above are featured by a single maximum and vanish at the boundaries $r\to \pm\infty$, we adopt the following outgoing wave boundary conditions
\bqn
\lb{Boundary}
R_B(r)&\sim&
\left\{
  \begin{array}{cc}
    e^{-i\frac{\omega}{c_T} r_*} & r\rightarrow-\infty\\
    e^{i\frac{\omega}{c_T} r_*}  & r\rightarrow\infty \\
  \end{array}
\right.\nb\\
R_C(r)&\sim&
\left\{
  \begin{array}{cc}
    e^{-i\frac{\omega}{c_V} r_*} & r\rightarrow-\infty\\
    e^{i\frac{\omega}{c_V} r_*}  & r\rightarrow\infty ,\\
  \end{array}
\right.
\eqn
where the tortoise coordinate $r_*=\int dr/f(r)$ for $m\ne 0$ and $r_*=r$ for $m=0$.

As discussed above, for the specific choice of model parameters, the master equations are decoupled, and the quasinormal frequencies can be calculated using the WKB method.
In this case, one is expected to find two independent spectra of quasinormal modes.
For coupled master equations, it is similar to the scenario where some interaction is introduced into a system of two damped harmonic oscillators.
The effect of the coupling in the master equation is an interesting subject and will be explored further.
Nonetheless, technically, when the system of coupled equations cannot be {\it diagonalized}, it poses a rather challenging task.
In the following section, we show that such coupled master equations can be solved using the finite difference method with reasonable precision by explicitly showing that both degrees of freedom attain identical frequencies.

\section{The quasinormal frequencies and their dependence on the coupling} \label{section4}
\renewcommand{\theequation}{4.\arabic{equation}} \setcounter{equation}{0}

In this section, we present the numerical results on the quasinormal modes by solving the master equations derived in the last section.
We will elaborate on the results of the quasinormal frequencies of both the decouple and coupled master equations, the late time tails, and the effect of coupling between the two degrees of freedom in the axial perturbations.

First, for the case $m=0$, the obtained quasinormal frequencies obtained by using the third order WKB method are given in Tab.~\ref{TableI}.
As the master equations are decoupled, the axial perturbations of the metric and {\ae}ther field give rise to two independent quasinormal spectra.
Since the relevant frequency scales with $c_T$ and $c_V$, the results are presented in terms of the ratios $\omega/c_T$ and $\omega/c_V$, respectively.
It is observed that for both spectra, at a given angular momentum $L$, the real part of the quasinormal modes largely remains unchanged, while the magnitude of the imaginary part grows with increasing overtone number $\bar{n}$.
For a given overtone number $\bar{n}$, the real part of the quasinormal modes increases with angular momentum, while the imaginary part mostly stays the same, in accordance with the eikonal limit~\cite{agr-qnm-geometric-optics-02}.
While the two spectra show similar properties, the magnitudes of the real and imaginary parts of $\omega/c_V$ are slightly larger than those of $\omega/c_T$ for the modes with identical quantum numbers.

The time profile of the axial perturbations of the metric and {\ae}ther field can be evaluated using the finite difference method.
To be specific, we rewrite the master equations as
\begin{equation}\lb{masterequation1B}
\begin{split}
&\frac{\partial^2R_{B, C}}{\partial u\partial v}+\frac{1}{4}V_{T, V} R_{B, C}+\frac{1}{4}U_{T, V} R_{C, B}=0 ,
\end{split}
\end{equation}
where one has restored the time partial derivative $\omega^2 \to -\frac{\partial^2}{\partial t^2}$ in Eqs.~\eqref{masterequation1} and utilized Eddington-Finkelstein coordinates $u=c_{T,V}t-r_*, v=c_{T,V}t+r_*$.
Subsequently, by carrying out the discretization process, the field on the grid sites can be computed according to
\begin{equation}\lb{masterequation1C}
\begin{split}
\tilde R_{B, C}^{i+1,j+1}=&\tilde R_{B, C}^{i-1,j+1}+\tilde R_{B, C}^{i+1,j-1}-\tilde R_{B, C}^{i-1,j-1}\\
&-\frac{\Delta v \Delta u}{8}\left[\tilde V^{i-1,j-1}_{T,V}\left(\tilde R_{B, C}^{i-1,j+1}+\tilde R_{B, C}^{i+1,j-1}\right)\right.\\
&\left.+\tilde U^{i-1,j-1}_{V,T}\left(\tilde R_{C, B}^{i-1,j+1}+\tilde R_{C, B}^{i+1,j-1}\right)\right] ,
\end{split}
\end{equation}
where we have assumed $c_V=c_T$ and made use of the notations
\begin{equation}
\begin{split}
R_{B, C}(t, r_*)&\equiv \tilde R_{B, C}\left(u, v\right)=\tilde R_{B, C}(i\Delta u, j\Delta v)=\tilde R_{B, C}^{i,j}, \\
V_{T, V}(r(r_*))&\equiv \tilde V_{T, V}\left(u, v\right)=\tilde V_{T, V}^{i, j}, \\
U_{T, V}(r(r_*))&=\tilde U_{T, V}^{i, j} .
\end{split}
\end{equation}

The numerical results are shown in FIG.~\ref{Fig1}.
As discussed below, the quasinormal frequencies can be extracted using the Prony method~\cite{agr-qnm-55, agr-qnm-lq-02} and are found to be consistent with those obtained above using the WKB approach.
For instance, for $L=3$, the two most dominate extracted quasinormal frequencies are $\omega_1 = 2.9554 - i 0.408754$ and $\omega_2 = 3.39952 - i 0.486151$.
For $L=4$, the values extracted from FIG.~\ref{Fig1} using the Prony methods are $\omega_1 = 4.08383 - i 0.446548$ and $\omega_2 = 4.42744 - i 0.489449$.
Both are readily compared with those given in Tab.~\ref{TableI}.

Moreover, the late-time tails are also present.
Although they are different from those in the Schwarzchild black hole, the asymptotical forms can be readily understood in terms of the specific power-law forms of the respective effective potentials.
In particular, according to the second line of Eqs.~\eqref{masterequation3}, the effective potential for {\ae}ther field gives $V_{\mathrm{eff}} = L(L+1)/r^2 + \bar{V}(r)$ as $r\to \infty$, where $\bar{V}(r) \sim r^{-4}$.
Numerically, for $L=1$ and $2$, it is verified that the late-time tails shown in FIG.~\ref{Fig1} are primarily gorverned by the form $t^{-(2L +4)}$.
Therefore, reminiscent of the scenarios for black holes, it is understood that the formation of the tails is due to the backscattering of the potential $\bar{V}(r)$ at spatial infinity~\cite{agr-qnm-tail-05, agr-qnm-tail-06}.
The latter gives rise to a branching cut on the negative part of the imaginary axis of the frequency-domain Green function, whose contribution is received chiefly in the vicinity of the origin.

For the case $m=1$ and $c_{14}=0$, as discussed above, the resulting master equation is also decoupled.
Therefore, one can also employ the WKB method to solve for the quasinormal frequencies.
The obtained quasinormal modes of axial perturbations are given in Tabs.~\ref{TableIIA} and~\ref{TableIIB}.
For the metric and {\ae}ther perturbations, it is found that for increasing overtone number $\bar{n}$ at a given angular momentum $L$, the real part of the quasinormal modes gradually decreases, while the magnitude of the imaginary part increases.
On the other hand, for a given overtone number $\bar{n}$, the real part of the quasinormal modes increases mainly linearly with angular momentum, and the imaginary part mostly remains the same.
The time profiles can be accessed by the finite difference method, and the results are shown in FIG.~\ref{Fig2}.
Also, the obtained results are consistent with those obtained by the WKB approach.
For instance, for the metric perturbations with $L = n = 2$, the two most dominate extracted quasinormal frequencies of $\omega/c_T$ are $\omega_1 = 0.295301 - 0.0692264 i $ and $\omega_2 = 0.259695 - 0.177727 i$.
For $L = 3$ and $n = 2$, the two most relevant modes are $\omega_1 = 0.499522 - 0.0763032 i$ and $\omega_2 = 0.460751 - 0.232638 i$.
Besides, the precision of these results encourages us to utilize the approach to extract the quasinormal frequencies for the scenario with nonvanishing coupling.

Now we turn to the case where the system of master equations Eqs.~\eqref{masterequation1}-\eqref{masterequation2} is generically coupled.
In this case, the two master equations are solved iteratively using Eqs.~\eqref{masterequation1C}.
The two oscillators evolve in time through the coupling and eventually reach a common eigenvalue, namely, the quasinormal frequency of the coupled system of master equations.
The time profiles obtained numerically for the case where $m=1$ and $c_{14} = 0.1$ and $3.0$ are presented in FIG.~\ref{Fig3}.
By employing the Prony method to extract the values of quasinormal frequencies, again, the results' robustness are readily verified.
For the axial metric oscillations, the two most dominant frequencies are extracted for three different time intervals, namely, $(170, 220)$, $(240, 290)$, and $(200, 500)$ as given in Tab.~\ref{TableIII}.
As expected, it is observed that resulting complex frequencies are mainly identical in value for both degrees of freedom, which indicates that they are indeed ``synchronized''.
Numerically, the values obtained using the Prony method are found reliable up to five significant figures, which are also invariant concerning either the interval of the fitting or the grid size.
In particular, the interval $(170, 220)$ was chosen because both the oscillation periods and magnitude variations of the two fields are visually different.
However, contrary to one's instinct, from Tab.~\ref{TableIII}, it is observed that the two extracted complex frequencies are almost identical, in agreement with the remaining results.
Such a dilemma can be understood by observing the specific values of the frequencies for the two foremost modes ($\bar{n}=0$ and $1$).
The real parts of frequencies are numerically close, and moreover, they are an order of magnitude larger than the imaginary parts of the frequencies.
As a result, a combination of them might give rise to {\it beat}.
The above justification can be confirmed by inspecting the weights of individual modes.
In the case $c_{14}=3.0$, for the metric perturbations, the respective weights of the two most dominant modes are $1.66281\times 10^{-6}$ and $3.47009 \times 10^{-7}$, and therefore, only the fundamental mode is practically observable.
For the {\ae}ther perturbations, on the other hand, the weights of the two modes are found to be $7.95414\times 10^{-7}$ and $5.15659\times 10^{-7}$, which are rather similar in magnitude.
As a result, the beat is observed in the early stage of the time profile of the {\ae}ther field, which can be observed in the top right plot of Fig.~\ref{Fig3}.
Also, we note that the extracted values for the modes of higher overtones are not as reliable as those of the fundamental mode.
In Tab.~\ref{TableIII}, we only present the results of the two foremost modes with $\bar{n} =0$ and $1$.

An interesting result is about the spectrum of quasinormal modes as a function of the coupling, notably the merger of two independent spectra into a unique one due to the presence of the coupling.
To be specific, the fundamental mode of the axial metric and {\ae}ther perturbations becomes the fundamental and first overtone modes of the coupled system.
This can be confirmed by observing the values in the columns of $c_{14}=0$ and $c_{14}$, respectively, in Tab.~\ref{TableIII}.
As the coupling grows, the real part of the fundamental mode decreases, while the magnitude of the imaginary part increases.
For the first overtone, both the real and magnitude of the imaginary parts of frequencies increase with increasing coupling.

\section{Further discussions and concluding remarks} \label{section5}
\renewcommand{\theequation}{4.\arabic{equation}} \setcounter{equation}{0}

In the present work, the Ellis drainhole solution is derived in Einstein-{\AE}ther gravity.
The obtained metric solution is asymptotically flat for both regions separated by the drainhole.
In Ref.~\cite{agr-einstein-aether-08}, it was pointed out that a static solution in vacuum Einstein-{\AE}ther gravity cannot be both regular and asymptotically flat.
This is consistent with the metric obtained in this study, where the scalar field $\Phi$ serves the role of the matter field.
Besides, from the wormhole perspective, the scalar field holds the throat open.

The quasinormal modes of the resulting drainhole are investigated by introducing the axial gravitational perturbations.
It is found that the derived master equations for the axial perturbations are featured by two coupled vector degrees of freedom.
Since the unperturbed metric is invariant under spatial reflection, the coupled nature of the obtained master equation implies that the two degrees will not mix axial and polar modes.
Subsequently, the quasinormal modes are studied by utilizing the finite difference method and WKB approximation.
In particular, the complex frequencies extracted using the Prony method are consistent with the specific values obtained by the WKB method when the coupling is turned off.
Moreover, the effect of the coupling on the resultant quasinormal frequency is studied.

From a physical viewpoint, the situation of coupled master equations is reminiscent of the interaction introduced into a system of two damped harmonic oscillators.
If the two oscillators are identical, a small coupling will break the degeneracy, resulting in two slightly different branches of the spectrum.
This is precisely the scenario described by the perturbation theory of the eigenvalue problem in quantum mechanics.
However, if the nature of the two oscillators is somehow distinct, even an insignificant strength of interaction might give rise to a non-trivial outcome.
Such a scenario has demonstrated itself in a few systems, such as the strongly damped $w$-mode encountered in pulsating relativistic stars~\cite{agr-qnm-star-05, agr-qnm-star-07}, first pointed out by Kokkotas and Schutz.
The results obtained in the present study indicate that the fundamental modes of the two degrees of freedom constitute the two lowest-lying modes of the coupled system.
Therefore, the underlying physics is of the former type, where the coupling leads to a merger of the two initially independent spectra and deforms it continuously as its strength increases.

We have employed the finite difference and Prony methods to extract the quasinormal frequencies numerically.
For such coupled master equations, however, another seemly possible approach is to utlize the matrix method~\cite{agr-qnm-lq-matrix-01, agr-qnm-lq-matrix-02, agr-qnm-lq-matrix-04}.
To be specific, one may write down the system of master equations in terms of a matrix equation whose size is adapted to include both degrees of freedom, similar to the case for Kerr black holes~\cite{agr-qnm-lq-matrix-03}.
Unfortunately, the resultant algebraic equation turns out to be highly nonlinear, and its complex root does not converge straightforwardly.
Therefore, only the finite difference method has been employed in the numerical approach.
Even though the obtained numerical values in this work are reinforced by satisfactory precision, it would be desirable if another independent approach could verify the results.

Last but not least, the effective potential of the master equation is featured by a single maximum outside of the throat, as shown in FIG.~\ref{Fig4}.
If, however, the effective potential possesses a second local maximum in the spacetime on the other side of the throat, such as the Damour-Solodukhin wormhole~\cite{agr-wormhole-qnm-02, agr-qnm-echoes-16}, one might expect echoes in the temporal profiles.
This might be another interesting subject to be explored further.

\begin{acknowledgments}
We gratefully acknowledge the financial support from
National Natural Science Foundation of China (NNSFC) under contract No. 11805166.
We also acknowledge the financial support from
Funda\c{c}\~ao de Amparo \`a Pesquisa do Estado de S\~ao Paulo (FAPESP),
Funda\c{c}\~ao de Amparo \`a Pesquisa do Estado do Rio de Janeiro (FAPERJ),
Conselho Nacional de Desenvolvimento Cient\'{\i}fico e Tecnol\'ogico (CNPq),
Coordena\c{c}\~ao de Aperfei\c{c}oamento de Pessoal de N\'ivel Superior (CAPES).
A part of this work was developed under the project Institutos Nacionais de Ciências e Tecnologia - Física Nuclear e Aplicações (INCT/FNA) Proc. No. 464898/2014-5.
The numerical part of the research is also supported by the Center for Scientific Computing (NCC/GridUNESP) of the S\~ao Paulo State University (UNESP).
\end{acknowledgments}

\bibliographystyle{h-physrev}
\bibliography{references_qian}


\begin{widetext}

\begin{table}[ht]
\caption{\label{TableI}The quasinormal frequencies $\omega/c_T$ and $\omega/c_V$ for the metric parameter $m=0$ and $n=1$.
By employing the third-order WKB approximation, the calculations are carried out using different values of the overtone number $\bar{n}$ and angular momentum $L$.}
\begin{tabular}{c c c c c}
         \hline
$\bar{n}$&$L$&~~~$\omega/c_T$~~~&~~~$\omega/c_V$~~~
        \\
        \hline
$0$&$3$&$2.9689 - 0.4351 i$&$3.3893 - 0.4871 i$
          \\
$1$&&$3.0117 - 1.3947 i$&$3.2321 - 1.4821 i$
          \\
$2$&&$3.2259 - 2.5063 i$&$2.9414 - 2.5302 i$
          \\
        \hline
$0$&$4$&$4.0760 - 0.4504 i$&$4.4150 - 0.4922 i$
          \\
$1$&&$3.9908 - 1.3684 i$&$4.2970 - 1.4880 i$
          \\
$2$&&$3.8420 - 2.3284 i$&$4.0710 - 2.5146 i$
          \\
        \hline
\end{tabular}
\end{table}

\begin{table}[ht]
\caption{\label{TableIIA} The quasinormal frequencies $\omega/c_T$ of the metric perturbations.
By employing the third-order WKB approximation, the calculations are carried out using the metric parameters $m=1$ and $c_{14}=0$ for different values of $n$, overtone number $\bar{n}$ and angular momentum $L$.}
\begin{tabular}{c c c c c}
         \hline
$\bar{n}$&$n$&~~~$L=2$~~~&~~~$L=3$~~~&~~~$L=4$~~~
        \\
        \hline
$0$&~~$2$~~&$0.2950 - 0.0774 i$&$0.4988 - 0.0768 i$&$0.6842 - 0.0792 i$
          \\
$1$& &$0.2533 - 0.2489 i$&$0.4723 - 0.2337 i$&$0.6668 - 0.2394 i$
          \\
$2$& &$0.1971 - 0.4407 i$&$0.4233 - 0.3996 i$&$0.6337 - 0.4044 i$
          \\
        \hline
$0$&$3$&$0.2498 - 0.0669 i$&$0.4368 - 0.06459 i$&$0.6027 - 0.06864 i$
          \\
$1$&&$0.2072 - 0.2309 i$&$0.4080 - 0.1951 i$&$0.5858 - 0.2072 i$
          \\
$2$&&$0.1712 - 0.4200 i$&$0.3495 - 0.3316 i$&$0.5528 - 0.3493 i$
          \\
        \hline
$0$&$4$&$0.2137 - 0.0579 i$&$0.3855 - 0.0550 i$&$0.5334 - 0.0600 i$
          \\
$1$&&$0.1778 - 0.2204 i$&$0.3583 - 0.1636 i$&$0.5184 - 0.1810 i$
          \\
$2$&&$0.1767 - 0.4083 i$&$0.2972 - 0.2734 i$&$0.4888 - 0.3048 i$
          \\
        \hline
\end{tabular}
\end{table}

\begin{table}[ht]
\caption{\label{TableIIB} The quasinormal frequencies $\omega/c_V$ of the {\ae}ther perturbations.
By employing the third-order WKB approximation, the calculations are carried out using the metric parameters $m=1$ and $c_{14}=0$ for different values of $n$, overtone number $\bar{n}$ and angular momentum $L$.}
\begin{tabular}{c c c c c}
         \hline
$\bar{n}$&$n$&~~~$L=2$~~~&~~~$L=3$~~~&~~~$L=4$~~~
        \\
        \hline
$0$&~~$2$~~&$0.3874 - 0.0767 i$&$0.5606 - 0.0797 i$&$0.7307 - 0.0810 i$
          \\
$1$& &$0.3614 - 0.2365 i$&$0.5412 - 0.2421 i$&$0.7156 - 0.2447 i$
          \\
$2$& &$0.3188 - 0.4076 i$&$0.5061 - 0.4114 i$&$0.6873 - 0.4128 i$
          \\
        \hline
$0$&$3$&$0.3439 - 0.0690 i$&$0.4974 - 0.0711 i$&$0.6481 - 0.0720 i$
          \\
$1$&&$0.3172 - 0.2131 i$&$0.4782 - 0.2160 i$&$0.6332 - 0.2176 i$
          \\
$2$&&$0.2733 - 0.3689 i$&$0.4434 - 0.3676 i$&$0.6052 - 0.3673 i$
          \\
        \hline
$0$&$4$&$0.3053 - 0.0615 i$&$0.4417 - 0.0632 i$&$0.5754 - 0.0640 i$
          \\
$1$&&$0.2791 - 0.1903 i$&$0.4235 - 0.1923 i$&$0.5615 - 0.1934 i$
          \\
$2$&&$0.2357 - 0.3306 i$&$0.3903 - 0.3277 i$&$0.5351 - 0.3267 i$
          \\
        \hline
\end{tabular}
\end{table}

\begin{table}[ht]
\caption{\label{TableIII} The quasinormal frequencies $\omega/c_T$ and $\omega/c_V$ for different coupling $c_{14}$ with the metric parameters $m=1$, $n=L=5$, and $c_T=c_V=1/\sqrt{1-c_+}$.
By employing the finite difference method and Prony method successively, the extracted complex frequencies are given for different fitting time intervals.
The results obtained by using the third-order WKB approximation is also presented for the case when $c_{14}=0$.}
\begin{tabular}{c c c c c}
         \hline
$c_{14}$&~~~interval~~~&$\bar{n}$&~~~$\omega/c_T$~~~&~~~$\omega/c_V$~~~
        \\
        \hline
$0$&$(200, 550)$&$0$&$0.602222 - 0.0551112 i$&$0.633124 - 0.0577419 i$
          \\
          &~~~WKB~~~&$0$&$0.60203 - 0.0550867 i$&$0.633 - 0.05761 i$
          \\
          &       &$1$&$0.591688 - 0.165982 i$&$0.622623 - 0.173659 i$
          \\
        \hline
$0.1$&$(170, 220)$&$0$&$0.601457 - 0.055102 i$&$0.601458 - 0.055104 i $
          \\
          &       &$1$&$0.633949 - 0.0575904 i $&$0.633932 - 0.0577492 i$
          \\
          &$(240, 290)$&$0$&$0.601458 - 0.0551037 i$&$0.601458 - 0.055104 i$
          \\
          &       &$1$&$0.633878 - 0.057765 i$&$0.633932 - 0.0577492 i$
          \\
          &$(200, 550)$&$0$&$0.601456 - 0.055105 i$&$0.601458 - 0.0551044 i$
          \\
          &       &$1$&$0.633882 - 0.0578415 i$&$0.633932 - 0.0577488 i$
          \\
        \hline
$1.0$&$(200, 550)$&$0$&$0.595801 - 0.0549862 i$&$0.595806 - 0.0549913 i$
          \\
          &       &$1$&$0.640019 - 0.0579054 i$&$0.63999 - 0.0578694 i$
          \\
        \hline
$2.0$&$(200, 550)$&$0$&$0.590965 - 0.0548367 i$&$0.590961 - 0.0548389 i$
          \\
          &       &$1$&$0.645282 - 0.0580371 i$&$0.645279 - 0.0580211 i$
          \\
        \hline
$3.0$&$(170, 220)$&$0$&$0.586927 - 0.0546892 i$&$0.586927 - 0.0546895 i$
          \\
          &       &$1$&$0.649758 - 0.0581647 i $&$0.649757 - 0.0581671 i$
          \\
          &$(240, 290)$&$0$&$0.586927 - 0.0546895 i$&$0.586927 - 0.0546892 i $
          \\
          &       &$1$&$0.649757 - 0.0581684 i$&$0.649756 - 0.0581664 i$
          \\
          &$(200, 550)$&$0$&$0.586927 - 0.0546897 i$&$0.586927 - 0.0546897 i$
          \\
          &       &$1$&$0.649757 - 0.0581673 i$&$0.649757 - 0.0581673 i$
          \\
        \hline
\end{tabular}
\end{table}

\end{widetext}

\begin{figure}[tbp]
\centering
\includegraphics[width=1\columnwidth]{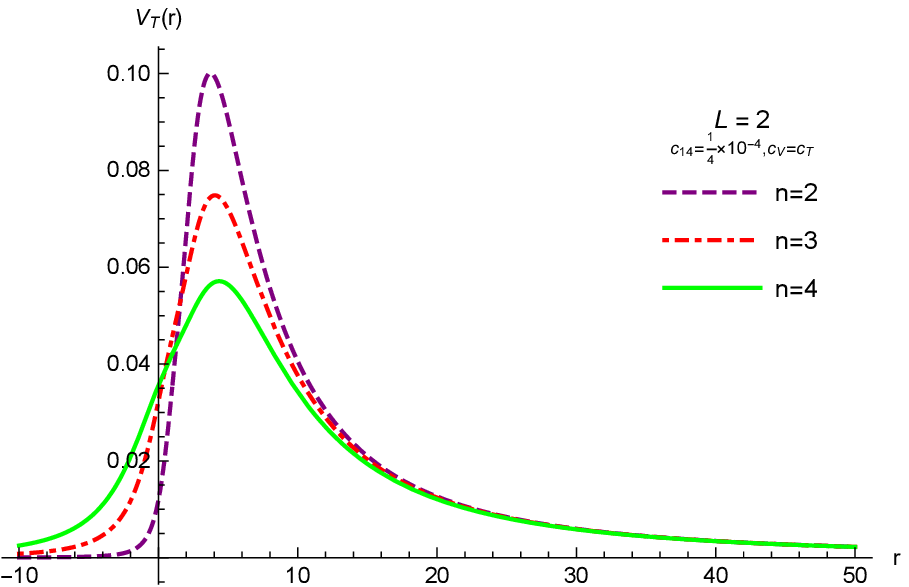}\includegraphics[width=1\columnwidth]{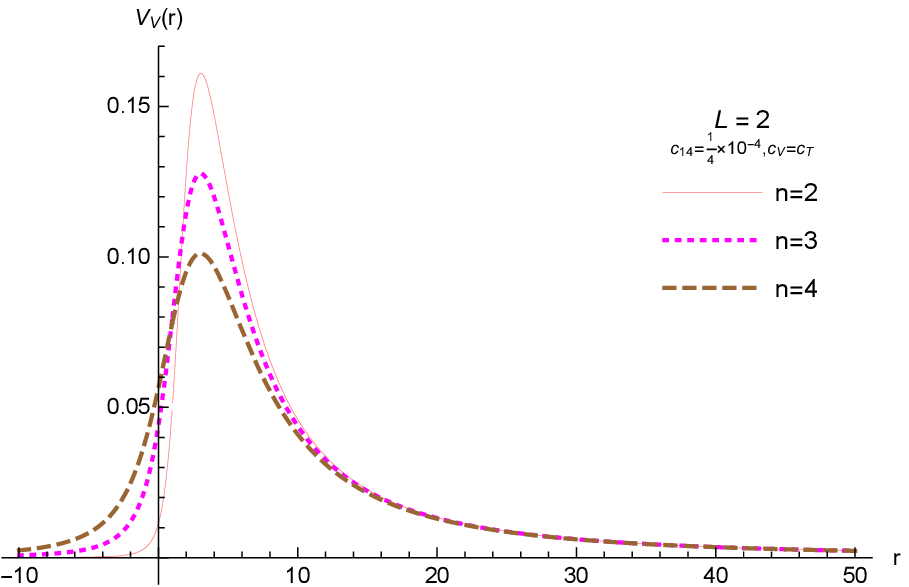}
\caption{The effective potentials $V_T$ and $V_V$ of the axial gravitational and {\ae}ther perturbations.}
\lb{Fig4}
\end{figure}

\begin{figure}[tbp]
\centering
\includegraphics[width=1\columnwidth]{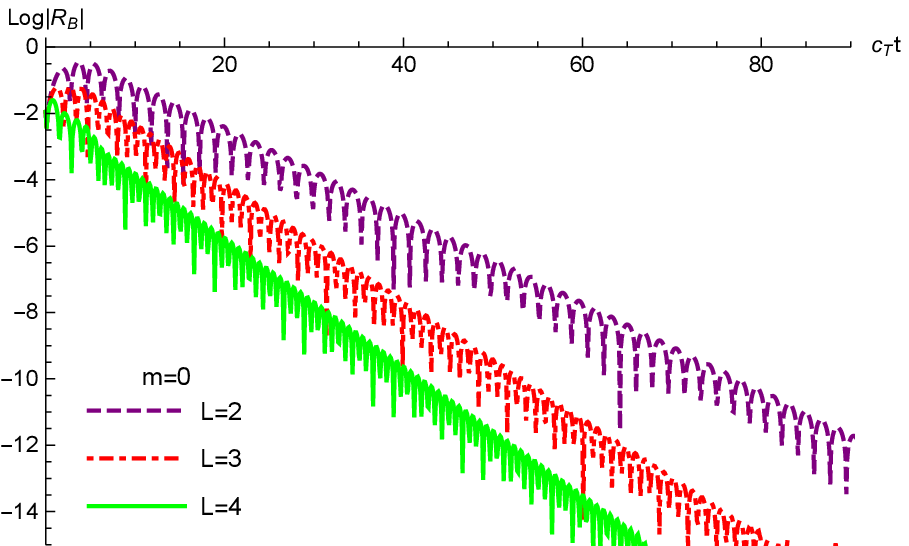}\includegraphics[width=1\columnwidth]{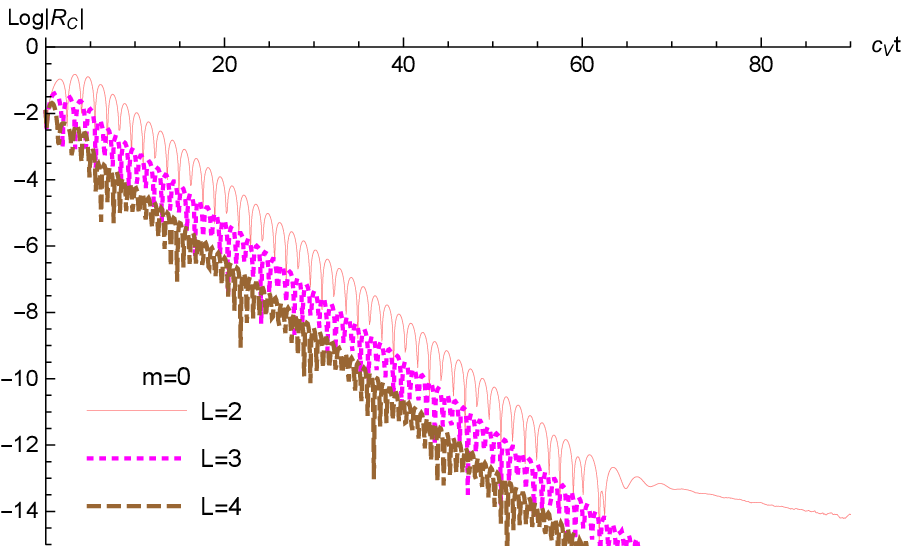}
\includegraphics[width=1\columnwidth]{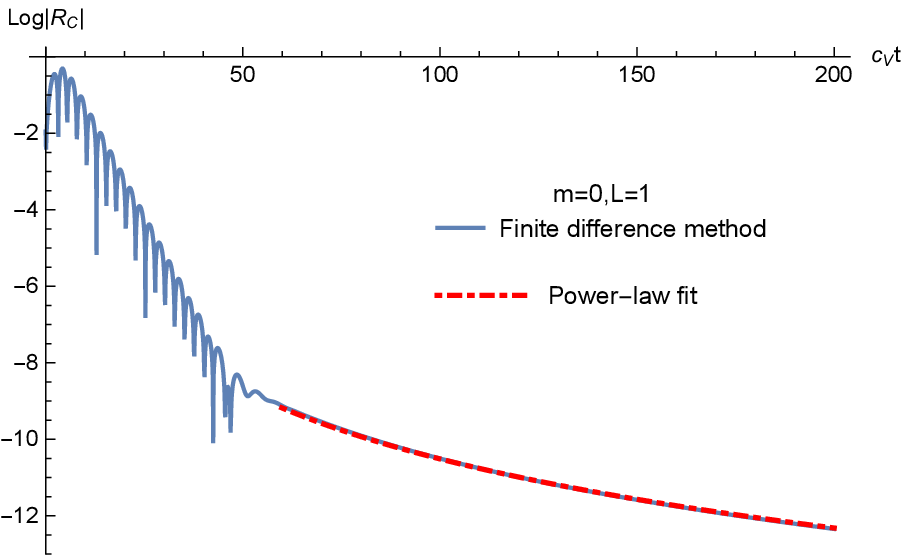}\includegraphics[width=1\columnwidth]{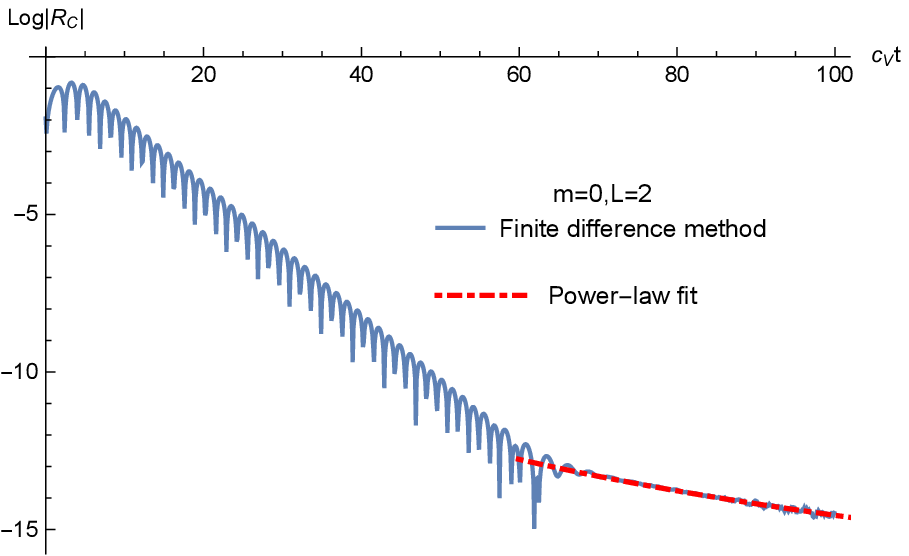}
\caption{The time profiles of axial gravitational quasinormal modes for $R_B$ and $R_C$ for the metric parameter $m=0$.
Bottom row: Close-ups on the {\ae}ther late-time tails together with the numerical fits in accordance to the respective power-law forms $t^{-(2L + 4)}$.}
\lb{Fig1}
\end{figure}

\begin{figure}[tbp]
\centering
\includegraphics[width=1\columnwidth]{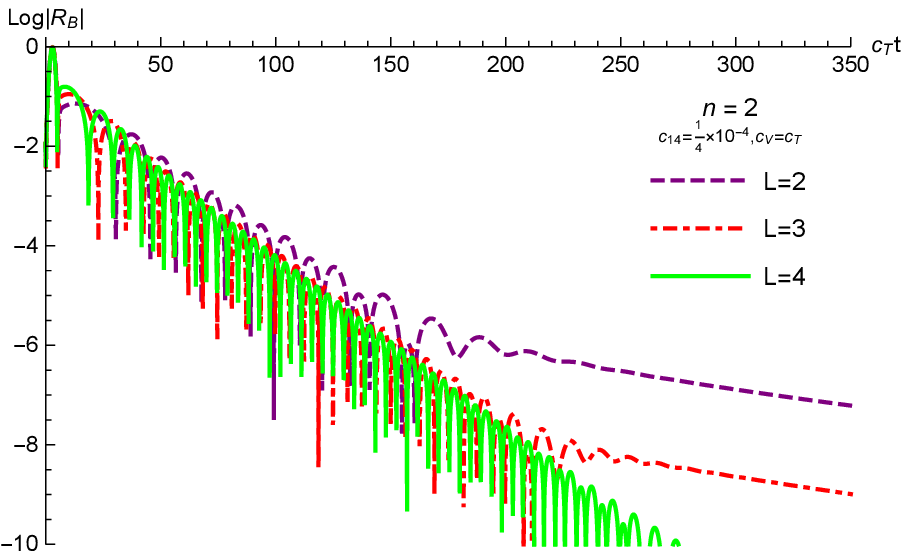}\includegraphics[width=1\columnwidth]{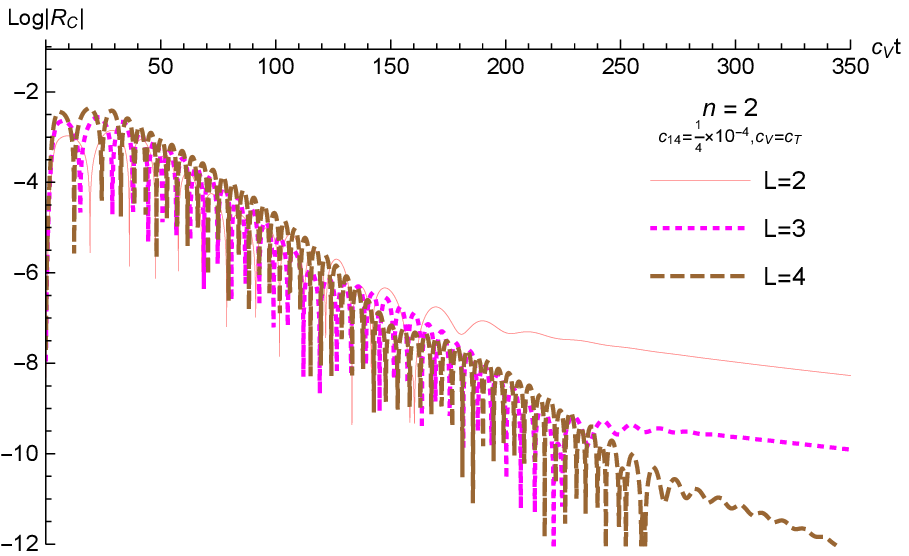}
\includegraphics[width=1\columnwidth]{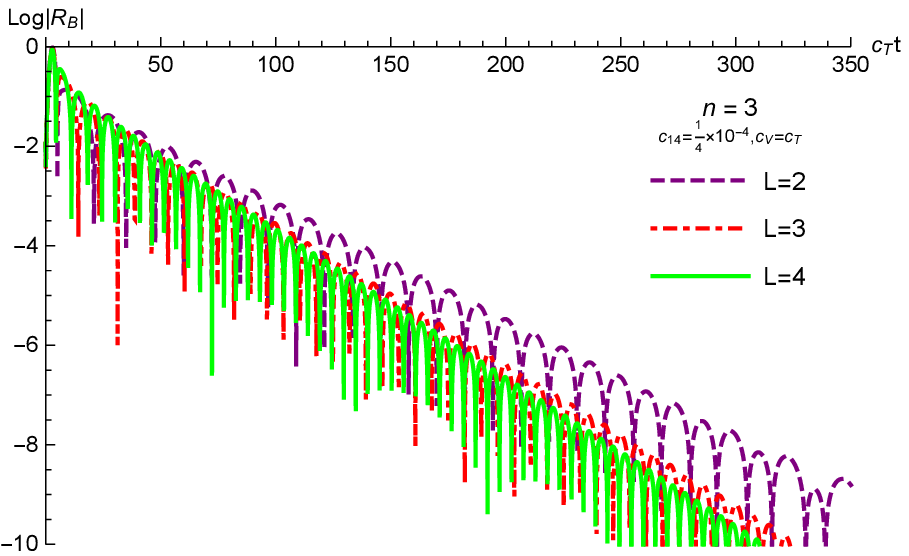}\includegraphics[width=1\columnwidth]{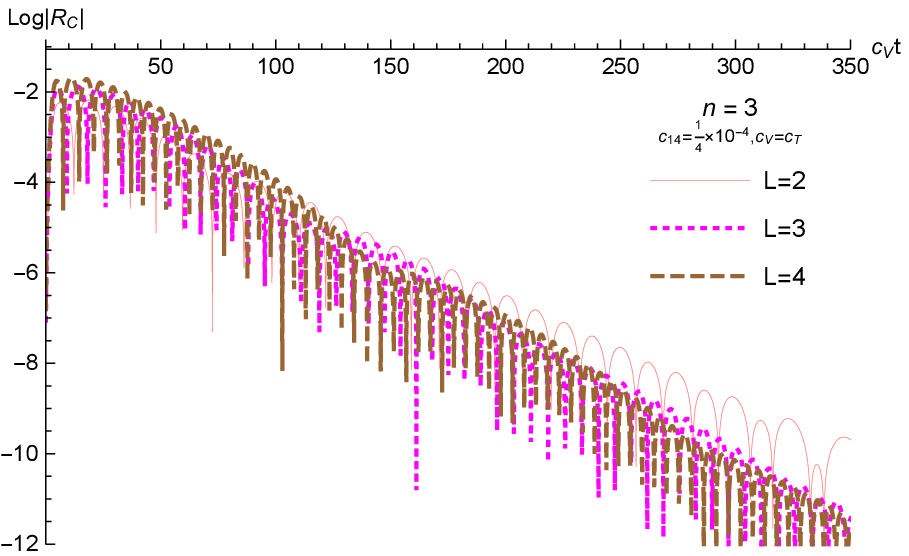}
\includegraphics[width=1\columnwidth]{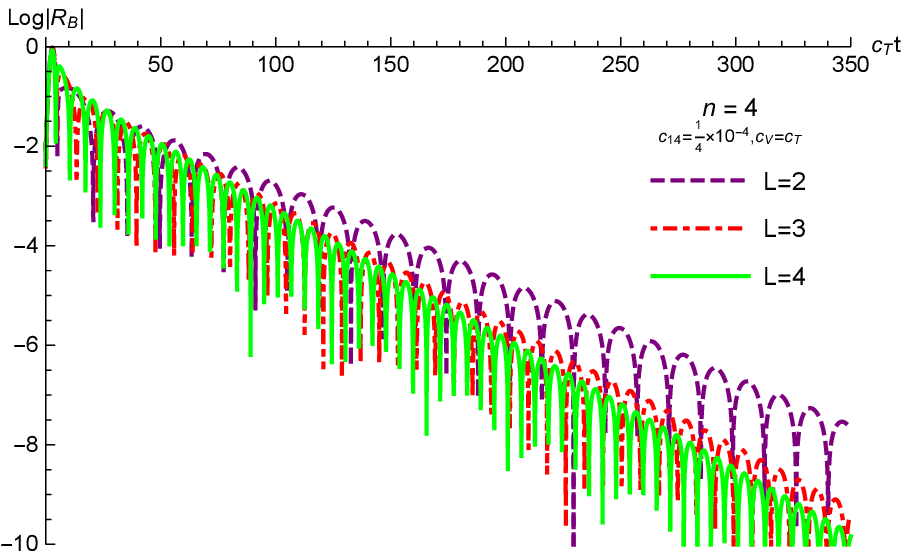}\includegraphics[width=1\columnwidth]{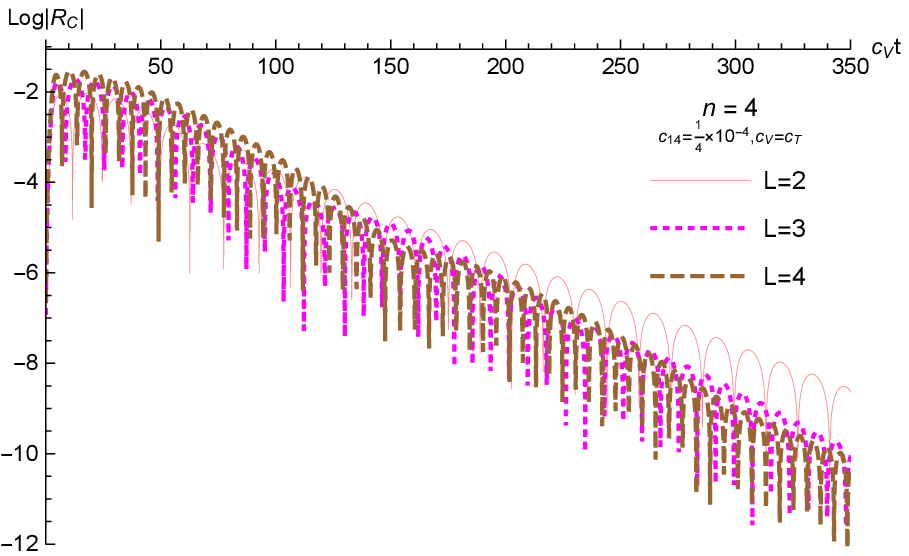}
\caption{The time profiles of axial gravitational quasinormal modes for $R_B$ and $R_C$ for the metric parameter $c_{14}=0$.}
\lb{Fig2}
\end{figure}

\begin{figure}[tbp]
\centering
\includegraphics[width=1\columnwidth]{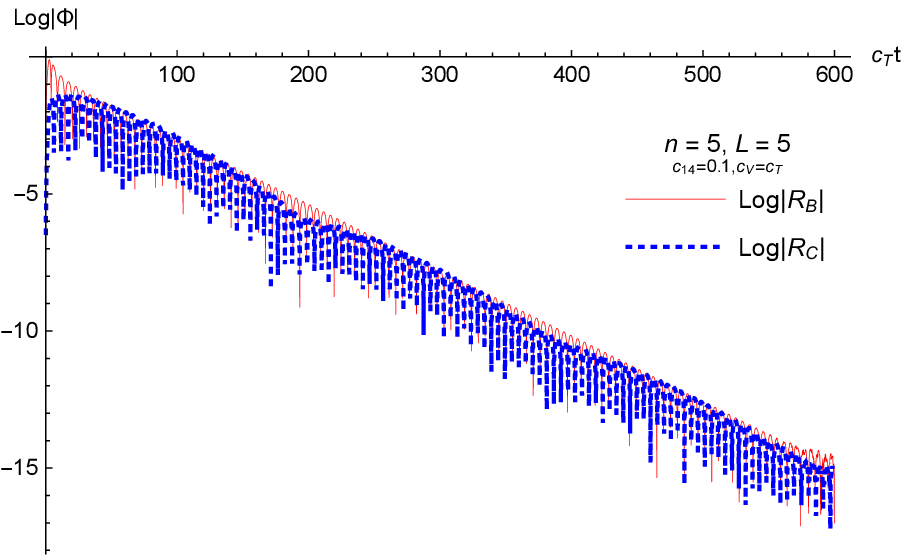}\includegraphics[width=1\columnwidth]{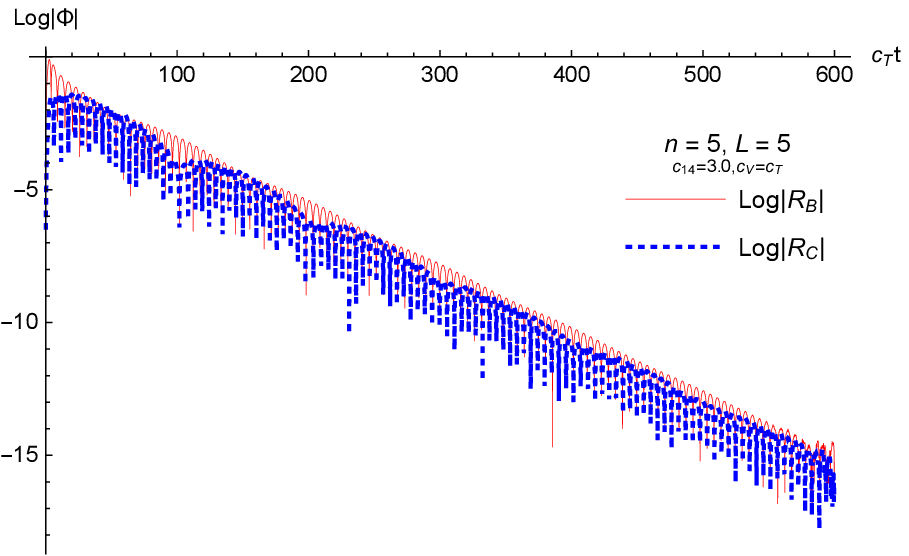}
\includegraphics[width=1\columnwidth]{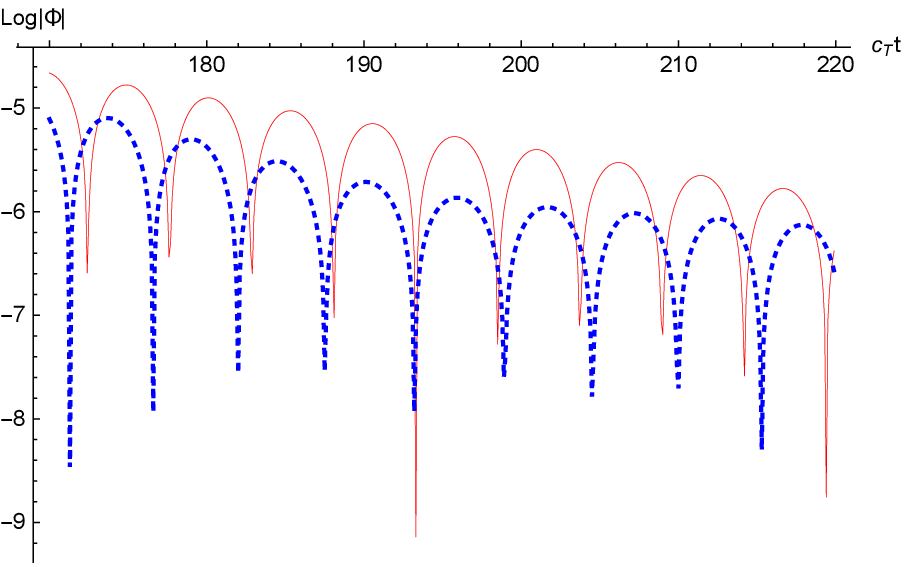}\includegraphics[width=1\columnwidth]{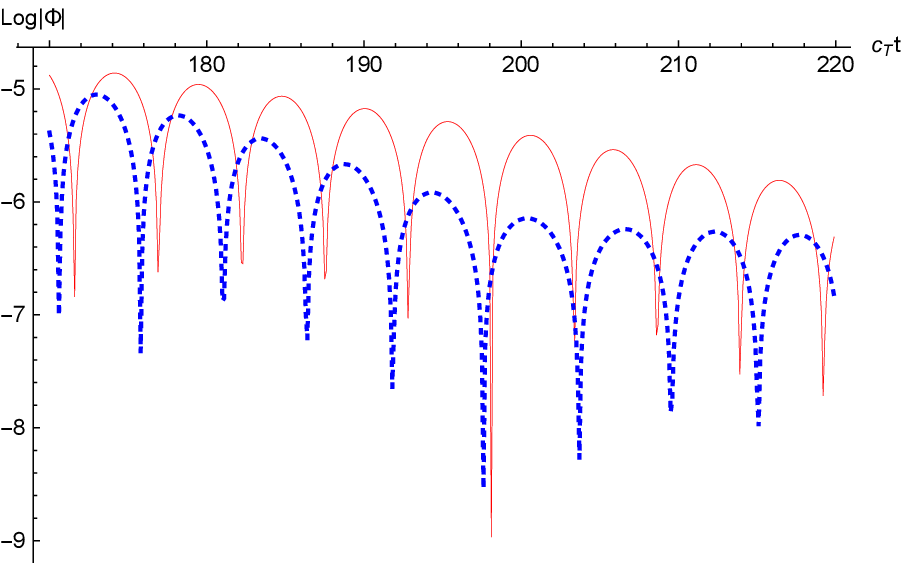}
\includegraphics[width=1\columnwidth]{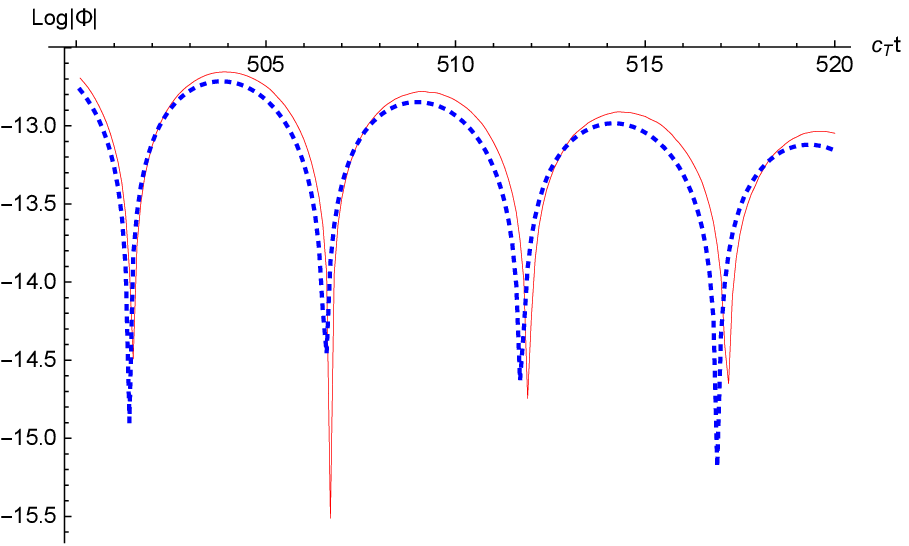}\includegraphics[width=1\columnwidth]{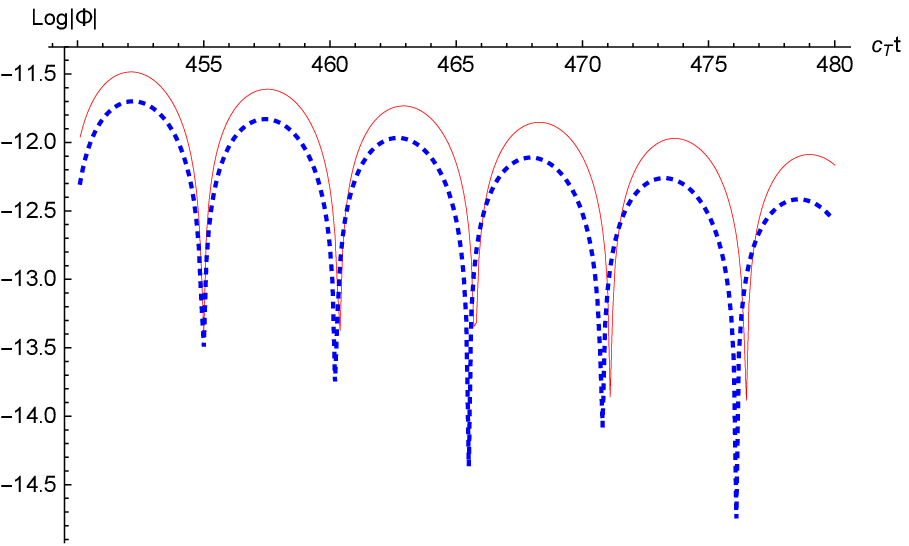}
\caption{The time profiles of axial gravitational quasinormal modes for $R_B$ and $R_C$ for the nonvanishing coupling $c_{14} \ne 0$.
The calculations are carried out using the parameters $m=1$, $L=n=5$, and $c_V = c_T = 1/(1-c_+)$.
The top left plot shows the results for a small coupling $c_{14} = 0.1$, while the two plots below it are close-ups that focus on different time intervals.
The right column is similar to the left column but calculations are done using a more significant coupling $c_{14} = 3.0$.}
\lb{Fig3}
\end{figure}

\end{document}